# Scalable deterministic integration of two quantum dots into an on-chip quantum circuit

*Shulun Li[1,2,3]†, Yuhui Yang[1]†, Johannes Schall[1], Martin von Helversen[1], Chirag Palekar[1], Hanqing Liu[2,3], Léo Roche[1], Sven Rodt[1], Haiqiao Ni[2,3], Yu Zhang[2,3], Zhichuan Niu[2,3]\*, Stephan Reitzenstein[1]\**

1. Institut für Festkörperphysik, Technische Universität Berlin, Hardenbergstraße 36, 10623 Berlin, Germany

2. State Key Laboratory for Superlattice and Microstructures, Institute of Semiconductors, Chinese Academy of Sciences, Beijing 100083, China

3. Center of Materials Science and Optoelectronics Engineering, University of Chinese Academy of Sciences, Beijing 100049, China

† S.L. and Y.Y. contributed equally to this paper.

ABSTRACT:

Integrated quantum photonic circuits (IQPCs) with deterministically integrated quantum emitters are critical elements for scalable quantum information applications and have attracted significant attention in recent years. However, scaling up them towards fully functional photonic circuits with multiple deterministically integrated quantum emitters to generate photonic input states remains a great challenge. In this work, we report on a monolithic prototype IQPC consisting of two pre-selected quantum dots deterministically integrated into nanobeam cavities at the input ports of a 2×2 multimode interference beam-splitter. The on-chip beam splitter exhibits a splitting ratio of nearly 50/50 and the integrated quantum emitters have high single-photon purity, enabling on-chip HBT experiments, depicting deterministic scalability. Overall, this marks a cornerstone toward scalable and fully-functional IQPCs.

Introduction:

Among different quantum emitter concepts, self-assembled III-V quantum dots (QDs) are of particular interest for future photonic quantum technologies. This is due to their almost ideal single-photon purity and indistinguishability, as well as their high spontaneous emission rate, which are crucial properties to meet the requirements of advanced quantum optics applications such as quantum communication, quantum computation, and quantum simulation[1-3]. Integrating QDs into nanophotonic structures boosts their photon extraction efficiency and improves their coherent properties by the Purcell effect[4-7]. Furthermore, monolithic integration of single QDs with passive components for quantum state generation and manipulation is an effective way to realize specific quantum tasks[8-11]. However, the random spectral and spatial distribution of self-assembled QDs hinders the scalable fabrication of complex integrated quantum photonic circuits (IQPCs) with multiple QDs generating the photonic input state.

To enable the controlled and scalable integration of single self-assembled QDs into photonic quantum devices, advanced nanofabrication technologies have been developed. For example, deterministic lithography techniques based on low-temperature optical imaging[12] and cathodoluminescence (CL) spectroscopy[13] have been used to integrate single QDs into micropillars[14], microlenses[15], circular Bragg gratings[16], and waveguide systems[11]. Here, we develop and use a marker-based scalable integration solution combining cryogenic CL spectroscopy and high-resolution electron beam lithography to deterministically embed two single InGaAs QDs into an IQPC based on GaAs/AlGaAs heterostructures. The IQPC consists of two nanobeam waveguides with integrated QDs which are connected to the input ports of an on-chip 2x2 multimode interferometer (MMI) beam splitter. In this system, single photons emitted by the QDs are coupled into the waveguides and transferred via the MMI section acting as 50/50 beam splitter to off-chip detectors for quantum optical measurements, as schematically shown in Fig. 1(a). We observed that both QDs showed multi-photon events of less than 5% in $g^{(2)}(\tau)$ and reasonable two-photon interference visibility under non-resonant excitation. Moreover, both QDs exhibit very strong multi-photon suppression determined via on-chip Hanbury Brown and Twiss (HBT) measurements.

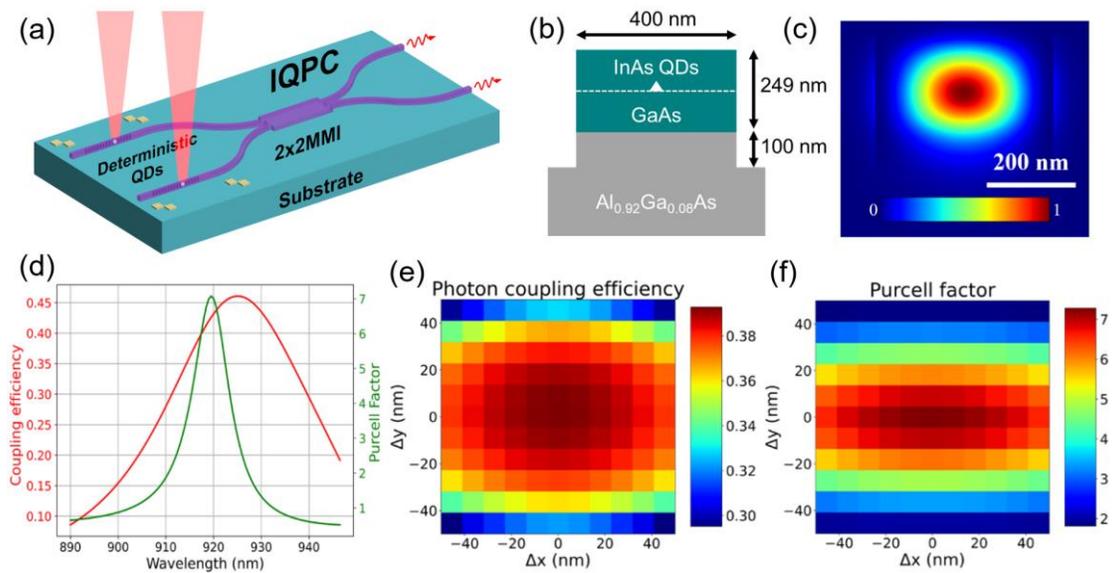

Fig. 1: (a) Schematic view of the IQPC device with two deterministically integrated QDs. (b) Layer design and cross-section of the investigated QD-waveguide heterostructure. (c) Fundamental TE mode of the ridge waveguide. (d) Numerically calculated photon coupling efficiency and Purcell factor of the nanobeam cavity. (e-f) 2D scan map of the calculated QD-waveguide coupling efficiency and Purcell factor under lateral displacement of the emitter with respect to the ideal position in the center of the nanobeam, where x (y) represents the parallel (perpendicular) direction to the WG axis.

To optimize the performance of the nanobeam cavity for high QD-waveguide coupling efficiency and Purcell factor, full 3D numerical simulations were performed using the finite element method (FEM) solver JCM suite. The QD emitter is modeled as a linear dipole source (polarized along the y-axis). The nanobeam cavity is designed such that its resonance wavelength corresponds to 920 nm, namely the central wavelength of the inhomogeneous broadened QD emission band. Nanobeam cavities optimized in this way yield a Purcell factor of approximately 7 and a QD-waveguide coupling efficiency $\eta$ >40% into the fundamental TE mode of the waveguide, as depicted in Fig1. (c). Thanks to the low-Q cavity (Q ~ 98), the nanobeam cavity exhibits broad-band Purcell enhancement with a FWHM of 9 nm, as shown in Fig. 1(d). Additionally, we studied the impact of fabrication imperfections, in terms of in-plane displacements of the emitter with respect to the perfect position in the center of the nanobeam, on the QD-waveguide coupling efficiency and on the Purcell factor. Fig. 1(e-f) display the obtained results as a function of the in-plane displacements of the dipole emitter ($\lambda$ = 920 nm) in a range of ±50 nm. The data shows that for a dipole displacement of up to 50 nm in both x- and y-direction, a coupling efficiency exceeding 29% and a moderate Purcell factor ~ 1.8 can be maintained.

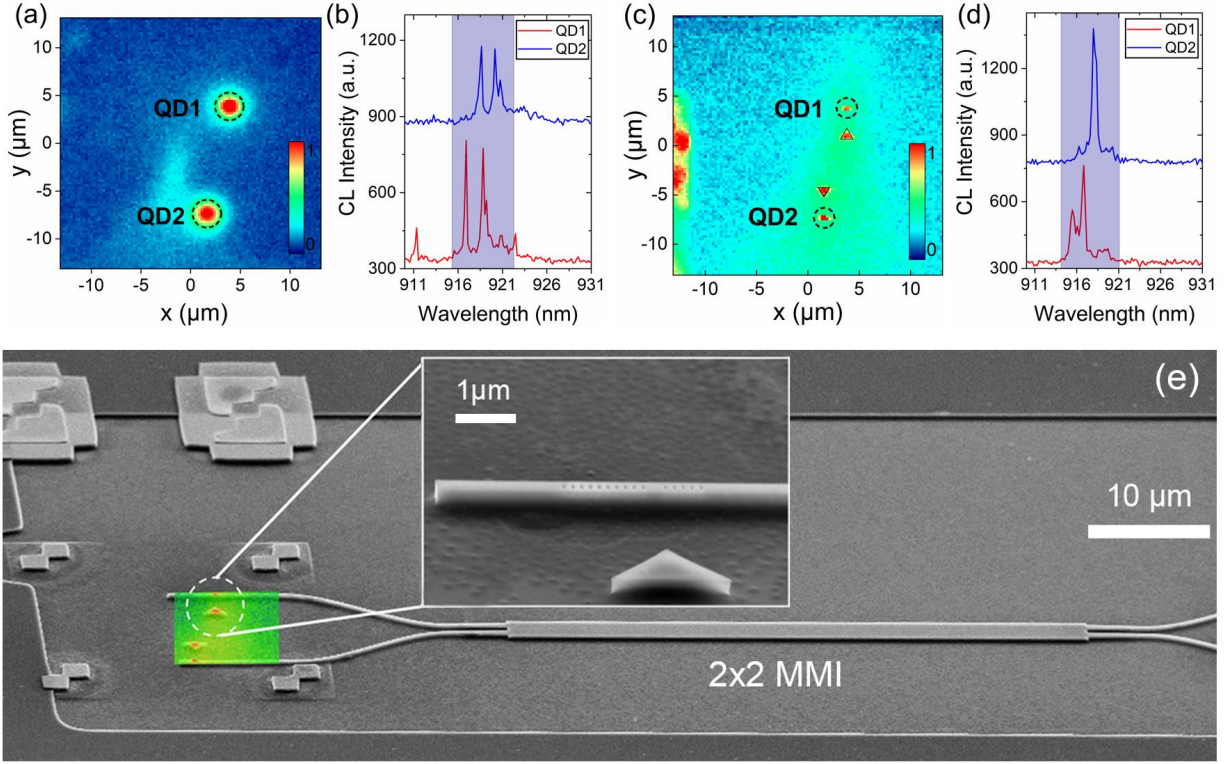

Fig. 2: (a) 2D CL map during the QD-selection process. Selected QDs (QD1 and QD2) are marked by black dashed circles. (b) CL spectra of the 2 QDs with similar emission wavelength close to the target value of 920 nm. (c) 2D CL map after fabrication. Two triangle indicators point out the target QD position in the center of the nanobeam cavity. (d) CL spectra of the two QDs taken after device fabrication to verify deterministic integration of the two selected QDs in the waveguide structures. (e) SEM image of the fabricated device. Inset: close-up view of the nanobeam cavity.

Device fabrication:

The QD-heterostructure sample was grown via molecular beam epitaxy (MBE) on a semi-insulating (100) GaAs substrate. It consists of a 243 nm thick GaAs core layer on top of a 1500 nm thick $Al_{0.9}Ga_{0.1}As$ cladding layer. The single layer of self-assembled In(Ga)As QDs is embedded in the core layer and was grown using the indium gradient growth method[17], which guarantees a low QD density of ~$10^7$ cm$^{-2}$ optimized for the fabrication of single-QD devices with emission wavelength ranging from 900 to 930 nm. To unify different coordinate systems during the marker-based deterministic device fabrication workflow, an array of metal square markers was patterned on the selected wafer area through an electron beam lithography (EBL) based lift-off process. Next, the sample was mounted onto the cold-finger of a Helium flow cryostat mounted to the interferometer stage of the state-of-the-art electron beam lithography system customized for deterministic device processing. CL maps were recorded at 20 K with an acceleration voltage of 20 keV and a pixel size of 250 nm x 250 nm to determine the lateral position and spectral features of QDs suitable for their device integration[18]. CL spectra of two selected QDs (QD1 and QD2) with a spatial separation of 15 μm and a spectral detuning of $\Delta\lambda = 1.73\ nm$ (for the dominant CL state peak) are presented in Fig. 2(a). The location of two QDs is precisely determined by applying a 2D Gaussian fitting function to the QD intensity distribution in the CL map with the spectral window of (918.65±1.35) nm, depicted by the blue area in Fig. 2(b). This fitting method gives a standard deviation of QD positions extraction accuracy of ±5-7 nm. The accurate coordinates of QDs also require precise knowledge of the marker position. Marker recognition is performed by using a home-made marker recognition script on the SEM image (pixel size of 12.5 nm × 12.5 nm) taken simultaneously with the CL mapping procedure. After acquiring the precise QD coordinates relative to the alignment markers, we performed EBL at room temperature to write the nanophotonic structures including ridge waveguides with a width of 400 nm, the two nanobeam cavities (10/5 holes), and a 2×2 MMI on-chip beam splitter section into negative-tone resist (CASR-6200) with a thickness of 330 nm. The center of the nanobeam cavity is aligned to the center of the QDs with an overall accuracy of (42±14) nm (for systematically evaluation, see *supporting information,* Fig. S6), including QD position extraction accuracy, marker recognition accuracy and EBL operation accuracy. To improve the sidewall smoothness and waveguide quality, we utilize a homemade proximity correction algorithm taking into account the specific properties of the used electron beam resist. Finally, inductively coupled plasma reactive ion etching (ICP-RIE) is applied to remove the unpatterned semiconductor material with an etching depth of 340 nm. Finalizing the fabrication workflow, successful deterministic QD integration and device processing are investigated by CL mapping (Fig. 2c-d). As shown in Fig. 2(e), the CL and SEM post-characterization images verify the successful integration of QD1 and QD2, as well as the excellent structural quality of the fabricated structures.

Optical properties:

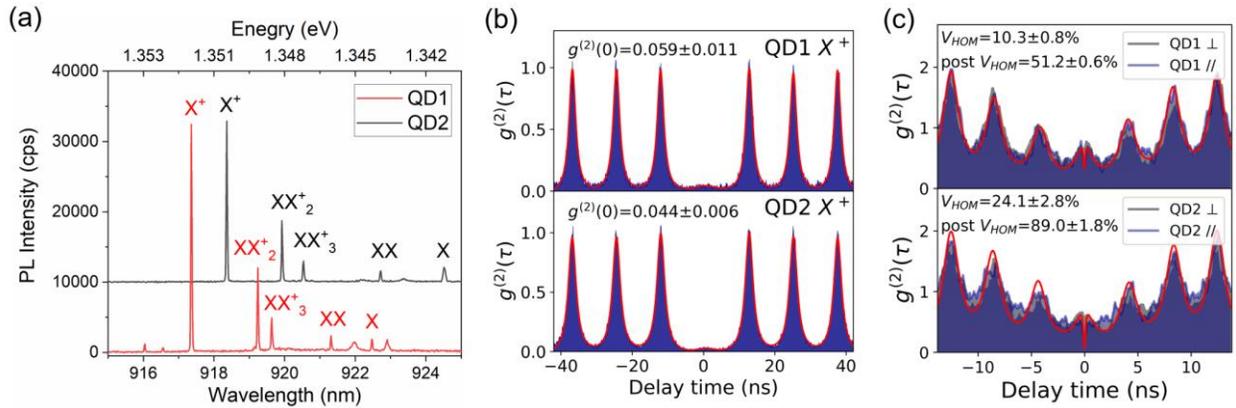

Fig. 3: Optical characterization of QD1 and QD2 deterministically integrated into the waveguide structures. (a) µPL spectra of the two QDs under wetting layer (865 nm) excitation at 4.17 K with tunable pulsed laser. (b) Second-order autocorrelation function of two QDs yielding a single-

photon purity of more than 94%. (c) Hong-Ou-Mandel interference experiments under pulsed wetting-layer excitation with 4 ns pulse delay, revealing a non-post-selected TPI visibilities of (10.3±0.8)% and (24.1±2.8)% and post-selected visibilities of (51.2±0.6)% and (89.0±1.8)% for QD1 and QD2 respectively.

Before performing measurements using on-chip routing of the photons by the waveguide structures, we first evaluate the optical properties of QD1 and QD2 by top excitation at the wetting layer wavelength (865 nm by a 2 ps-pulsed laser) and top collection in regular confocal µPL configuration. We identify the emission lines from each exciton state via power and polarization dependent µPL measurements (for details, see *supporting information*). The dominant line for both QDs originates from a positively charged exciton ($X^+$) with a spectral separation of 0.92 nm between QD1 and QD2 as shown in Fig. 3(a). Due to the different operating temperatures (4.7 K in µPL versus 20 K in CL) and the different excitation conditions, there is a 0.52 nm offset for the dominant peaks between the CL spectra and the µPL spectra while the overall spectral fingerprint of the QDs appears to be the same. The $X^+$ state is further filtered by the monochromator and sent to a fiber-based HBT setup including single-photon counting modules (SPCM) based on avalanche photo diodes with a time resolution of ~450 ps. As exhibited in Fig. 3(b), the second-order autocorrelation function at zero-time delay reveals a strong photon antibunching of $g^{(2)}(0) = 0.059 \pm 0.011$ and $g^{(2)}(0) = 0.044 \pm 0.006$ for QD1 and QD2, respectively. We further measure the two-photon interference (TPI) visibility from the aforementioned QDs using the superconducting nanowire single photon detectors (SNSPDs, time resolution: ~50 ps), The results in Fig. 3(c) give the non-postselected Hong-Ou-Mandel (HOM) visibility of (10.3±0.8)% and (24.1±2.8)% for QD1 and QD2 by comparing the area of center peak area with the area of the side peaks[19-21]. Fitting the HOM data using the equations based on Lorentzian peak functions: $f_\perp = \frac{2}{\pi}\left[A_0 \frac{\tau_1^2}{4(t-t_0)^2+\tau_1^2} + \sum_{i=1}^{3} A_i \frac{\tau_1^2}{4(t-t_i)^2+\tau_1^2} + \sum_{i=1}^{3} A'_i \frac{\tau_1^2}{4(t-t_i)^2+\tau_1^2}\right]$, $f_\parallel = \frac{2}{\pi}\left[A_0 \frac{\tau_1^2}{4(t-t_0)^2+\tau_1^2} - A_{00}\frac{\tau_c^2}{4(t-t_0)^2+\tau_c^2} + \sum_{i=1}^{3} A_i \frac{\tau_1^2}{4(t-t_i)^2+\tau_1^2} + \sum_{i=1}^{3} A'_i \frac{\tau_1^2}{4(t-t_i)^2+\tau_1^2}\right]$ allows us to determine the postselected two-photon visibility: $V_{HOM} = \frac{g_\perp^{(2)}(0)-g_\parallel^{(2)}(0)}{g_\perp^{(2)}(0)} = \frac{f_\perp(0)-f_\parallel(0)}{f_\perp(0)}$ of (51.2±0.6)% and (89.0±1.8)% for QD1 and QD2. We attribute the low non-postselected HOM visibilities to spectral noise introduced by surface charges in the proximity of the QDs with a distance to the etched surfaces of only ~50 nm[22].

Following the initial QD characterization, the sample is cleaved perpendicular to the waveguide orientation to expose the facet of the output waveguides for optical collection through a 50x objective lens (NA = 0.42) for lateral detection. To separate the µPL emission through the two output waveguides with a distance of 12 µm, we set up a telescope system composed of two lenses with focal lengths of 500 mm and 100 mm and a D-shaped mirror. The telescope system gives a magnification of 125× (see Fig. S3 in *supporting information* for details), which leads to a distance of the two outputs of 1.5 mm at the image plane, where the D-shape mirror is used to separate the signal from the two output facets of the waveguides. Using this µPL configuration, we first checked the splitting ratio of the 2×2 MMI by comparing the µPL spectra recorded from the two output facets, which yields a splitting ratio of (50.7 ± 5.1/49.3 ∓ 5.1)% (see Fig. 4(a-b)). Furthermore, we perform a polarization dependent measurement of waveguide-coupled emission of QD1 and QD2 to determine the degree of linear polarization (DOP) of the $X^+$ line. We obtained a linear DOP of 86% (90%) for QD1 (QD2) confirming that the dominant propagating mode in the waveguide is the fundamental TE mode.

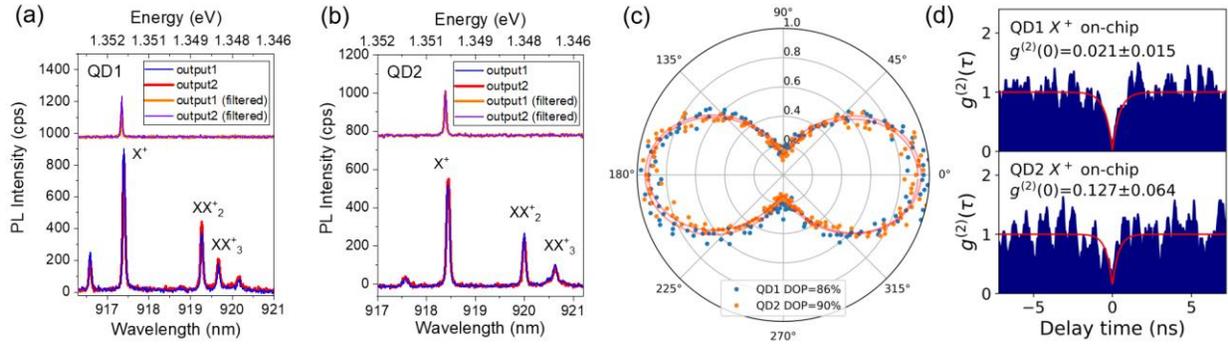

Fig. 4 (a-b): µPL spectra of QD1 and QD2 taken from two output facets of the waveguide circuit. A $(50.7 \pm 5.1/49.3 \mp 5.1)$% splitting ratio of the 2x2 MMI is observed over the whole spectrum. (c) Polarization dependent µPL measurements of QD1 and QD2 detected from the waveguide facets. Fitting the experimental data yields a degree of linear polarization of 86% (90%) for QD1 (QD2) for the device confirming light propagation though the dominant TE transmission mode in the waveguides. (d) On-chip HBT measurements via off-chip filtering and detection of each QD under 780 nm CW excitation, we obtained on-chip $g^{(2)}(0) = 0.021 \pm 0.015$ for QD1 and $g^{(2)}(0) = 0.127 \pm 0.064$ for QD2 respectively.

The overarching goal of integrated quantum photonic circuits is the scalable integration of quantum functionality in compact and robust chips[23]. To demonstrate the basic on-chip functionality of our deterministically fabricated QD-nanobeam-waveguide-MMI quantum photonic circuits, we performed HBT experiments by using the MMI as on-chip 50/50 beam splitter. Emission of the $X^+$ state detected from two output arms is filtered via two notch filters (in reflection configuration with a wavelength selection range of <0.2 nm) and coupled to two single mode fibers coupled to SPCMs for auto-correlation measurements. The normalized $g^{(2)}(\tau)$ of both integrated QDs recorded under CW off-resonant excitation is shown in Fig. 4(d). Without any deconvolution and background subtraction, we determine $g^{(2)}(0) = 0.021 \pm 0.015$ for QD1 and $g^{(2)}(0) = 0.127 \pm 0.064$ for QD2 by exponential fitting the experimental data, proving the quantum nature of emission and the scalability of deterministic device processing. We attribute the non-ideal multiphoton suppression to charge carrier recapture processes and uncorrelated background emission from further QDs in the waveguide and in the MMI section. This issue can be solved in the future in hybrid IQPC concepts which includes QDs only in the active GaAs section. It's worth noting that the different $g^{(2)}(0)$ values obtained from top-collection and on-chip measurements are mainly due to non-optimal coupling of signal as several loss channels on the off-chip detection path, e.g., poor coupling efficiency from waveguide facet and insertion loss of notch filter, which leads to lower signal-to-noise ratio. We note that strict resonant excitation might play a key role in reducing the carrier recapture and time jitter of single photons, which has been demonstrated to improve both single photon purity[24] and TPI visibility[25]. However, in the present case inappropriate signal-to-noise ratio obstructs the observation of resonance fluorescence under off-chip detection. In future optimizations we plan to address and solve both issues by including for instance grating outcoupling elements or tapered waveguides for better photon outcoupling efficiency, and to enable experiments under resonant excitation.

Conclusion and outlook:

In conclusion, we demonstrated the deterministic fabrication of two separated quantum dots into nanobeam cavities and 2x2 on-chip MMI into a monolithic IQPC. The overall alignment accuracy of $(42\pm14)$ nm between two similar QDs and the nanobeam waveguides is achieved through spectral and spatial preselection of QDs via CL mapping before marker-based EBL. The on-chip MMI section has a splitting ratio of $(50.7\pm5.1/49.3\mp5.1)\%$ and enables HBT experiments with an on-chip beam splitter, which shows clear evidence of quantum nature and chip-level scalability. Combining our scalable deterministic fabrication concept with local spectral fine-tuning of QDs using the quantum confined Stark effect[19] has high potential to lead to fully functional on-chip integrated quantum photonic circuits. Our deterministic fabrication technology and IQPC concept, in combination with on-chip phase-shifters and detectors[24] can pave the way for implementing advanced quantum functionality such as on-chip TPI[27] and fully on-chip boson sampling.

ASSOCIATED CONTENT

**Supporting Information**

Exciton states confirmation, Purcell enhancement evaluation, optical setup details and CW HOM results. (PDF)

AUTHOR INFORMATION

**Corresponding Author**

* Stephan Reitzenstein

E-Mail: stephan.reitzenstein@physik.tu-berlin.de

\* Zhichuan Niu

E-Mail: zcniu@semi.ac.cn

**Present Addresses**

1. Institut für Festkörperphysik, Technische Universität Berlin, Hardenbergstraße 36, 10623 Berlin, Germany

2. State Key Laboratory for Superlattice and Microstructures, Institute of Semiconductors, Chinese Academy of Sciences, Beijing 100083, China

3. Center of Materials Science and Optoelectronics Engineering, University of Chinese Academy of Sciences, Beijing 100049, China

**Author Contributions**

S.R. conceived the project. Y.Y. and L.R. performed the numerical simulations. S.L., H.L, and H.N. grew the QD samples. S.L., Y.Y., J.S. performed the CL experiments and electron beam lithography to deterministically fabricate the devices with the help of S.R. S.L., M.H. and C.P. built the set-up and characterized the devices. S.L., Y.Y., C.P. and Y.Z. analyzed the data. S.L., Y.Y. and S.R. wrote the manuscript with contributions from all authors. S.R. and Z.N. supervised the project.

**Funding Sources**

Funds used to support the research of the manuscript: (1) National Key Technologies R&D Program of China, 2018YFA0306100. (2) Key-Area Research and Development Program of Guangdong Province (Grant No. 2018B030329001). (3) National Natural Science Foundation of China, 62035017, 61505196. (4) European Union's Horizon 2020 Research and innovation Programme under the Marie Sklodowska-Curie Grant Agreement No. 861097 (QUDOT-TECH). (5) Einstein foundation via the Einstein Research Unit "Perspectives of a quantum digital transformation: Near-term quantum computational devices and quantum processors". (6) German Research Foundation (DFG) via grants RE2974/29-1 and INST 131/795-1 FUGG.

**ACKNOWLEDGMENT**

S.L. thanks the scholarship sponsored by UCAS to offer the opportunity to conduct this research. S.L. and Y.Y. thanks the mechanical workshop of TUB for the supportive help to fabricate a new design of the cryostat shield. We thank Ronny Schmidt and Stefan Bock for technical support in the nano-processing workflow. We thank Sven Burger for providing access to the JCM FEM solver. We thank the HPC staff of the math department in TU Berlin for helping and providing the usage of high-performance computer.

**ABBREVIATIONS**

IQPCs Integrated quantum photonic circuits; HBT Hanbury Brown and Twiss; MBE molecular beam epitaxy; EBL Electron Beam Lithography; SEM scanning electron microscope; QD quantum dot; CL cathodoluminescence; MMI multimode interferometer; HOM Hong-Ou-Mandel; OPO optical parameters oscillation; DOP degree of polarization.

**Present address:**

S.L.: Institute of Semiconductors, Chinese Academy of Sciences, Beijing, 100083, China

Y.Y.: Technische Universität Berlin, Berlin, 10623, Germany

J.S.: Technische Universität Berlin, Berlin, 10623, Germany

M.H.: Technische Universität Berlin, Berlin, 10623, Germany

C.P.: Technische Universität Berlin, Berlin, 10623, Germany

H.L.: Institute of Semiconductors, Chinese Academy of Sciences, Beijing, 100083, China

L.R.: Technische Universität Berlin, Berlin, 10623, Germany

S.R.: Technische Universität Berlin, Berlin, 10623, Germany

H.N.: Institute of Semiconductors, Chinese Academy of Sciences, Beijing, 100083, China

Y.Z.: Institute of Semiconductors, Chinese Academy of Sciences, Beijing, 100083, China

Z.N.: Institute of Semiconductors, Chinese Academy of Sciences, Beijing, 100083, China

S.R.: Technische Universität Berlin, Berlin, 10623, Germany

TOC Graphic

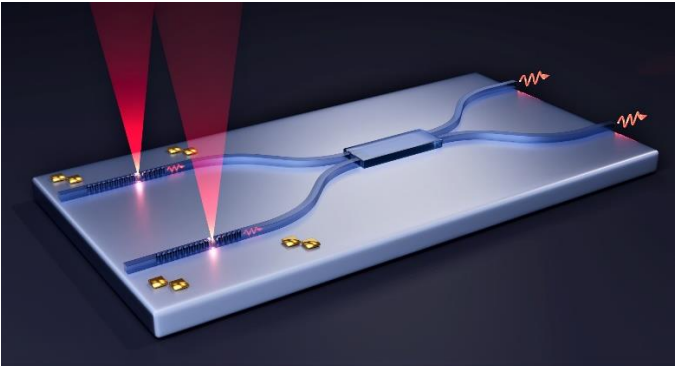

# Supporting Information of

*Scalable deterministic integration of two quantum dots into an on-chip quantum circuit*

*Shulun Li[1,2,3†], Yuhui Yang[1†], Johannes Schall[1], Martin von Helversen[1], Chirag Palekar[1], Hanqing Liu[2,3], Léo Roche[1], Sven Rodt[1], Haiqiao Ni[2,3], Yu Zhang[2,3], Zhichuan Niu[2,3*], Stephan Reitzenstein[1*]*

1. Institut für Festkörperphysik, Technische Universität Berlin, Hardenbergstraße 36, 10623 Berlin, Germany

2. State Key Laboratory for Superlattice and Microstructures, Institute of Semiconductors, Chinese Academy of Sciences, Beijing 100083, China

3. Center of Materials Science and Optoelectronics Engineering, University of Chinese Academy of Sciences, Beijing 100049, China

† S.L. and Y.Y. contributed equally to this paper.

**(i). Identification of exciton states of deterministically integrated QDs.**

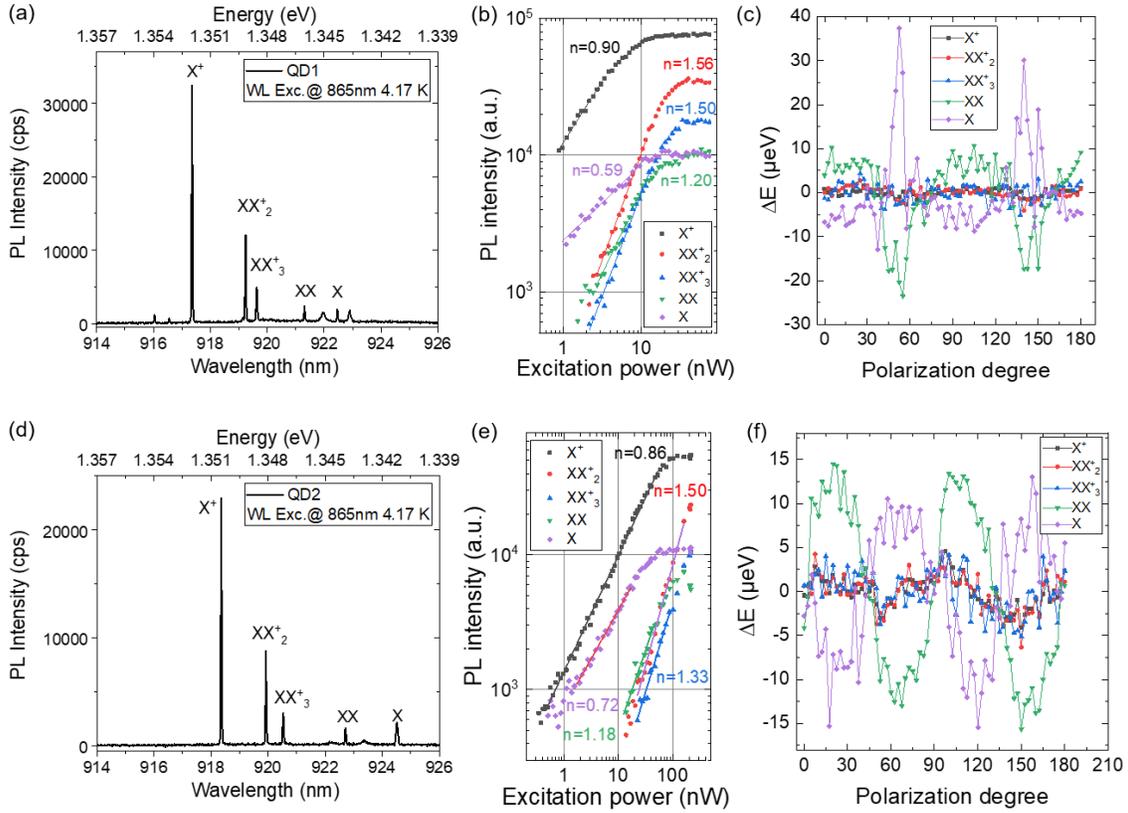

Fig. S1: Power and polarization dependent μPL measurements of 2 deterministically integrated QDs (QD1 and QD2). The obtained results allow one to identify the excitonic states of the QDs.

We performed power-, polarization- and time-dependent μPL studies to assign the emission lines to the exciton states of QD1 and QD2. QD1 and QD2 show similar spectral features. Analyzing the power-dependent scaling of the QD emission, single exciton states (X) and biexciton states (XX) were identified by comparing the scaling factor (i.e., the slope of the μPL intensity in log-log scale). The determined values are $0.59\pm0.03$ ($0.72\pm0.02$) for QD1 (QD2) X state and $1.20\pm0.03$ ($1.18\pm0.04$) for QD1 (QD2) XX state which leads to a ratio of $2.03\pm0.02$ and $1.64\pm0.02$ between the XX an X slope which is close to the theoretically expected value of 2. Additionally, charged excitonic states are identified by polarization dependent µPL studies which reveal an anti-correlated behavior of XX and X as can be seen in Fig. 1(c) and (f), while trion states do not show a significant polarization dependence.

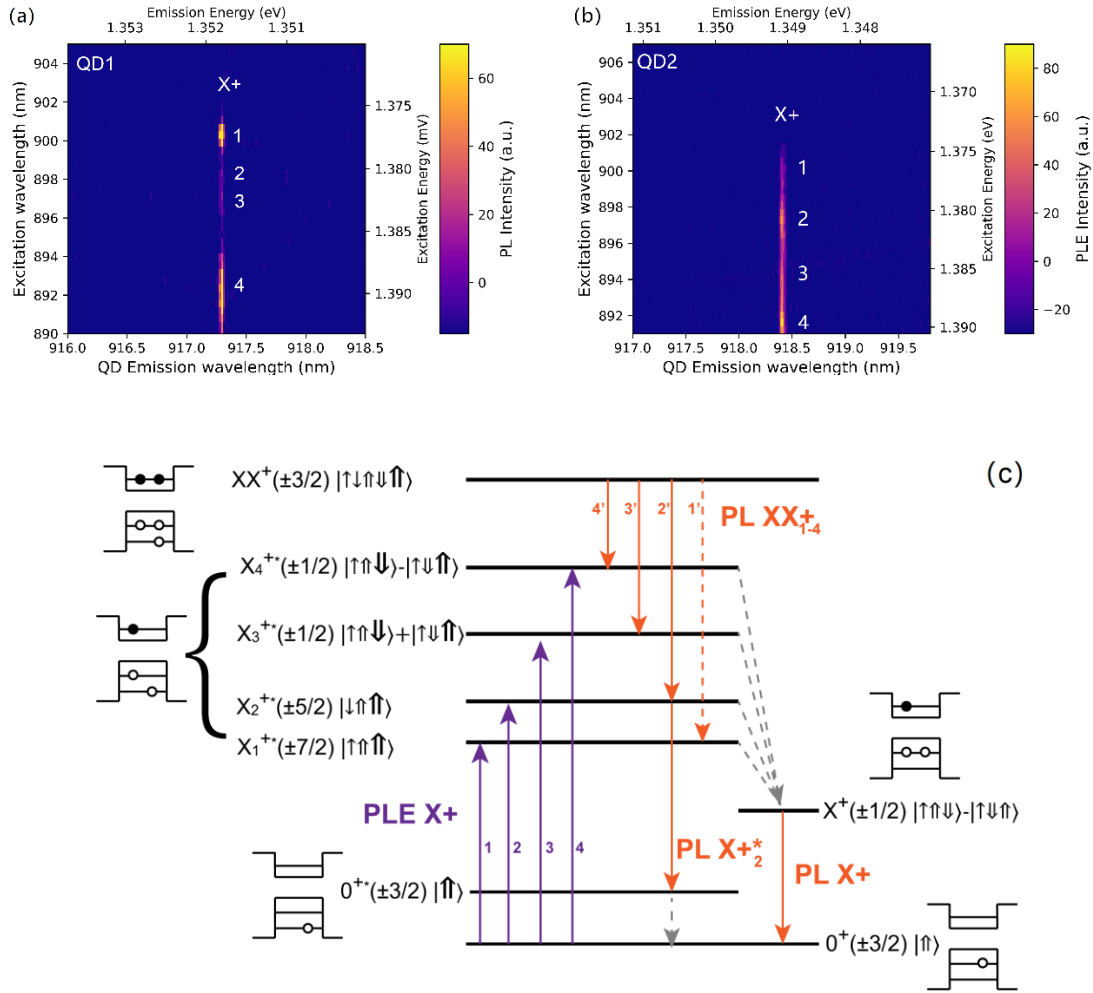

Fig. S2 (a-b): Micro-photoluminescence excitation (µPLE) spectra of 2 QDs. (c) Energy-level diagram sketch of positively charged biexciton (XX+) and excited positive trion (X+*). One of the possible spin configurations is indicated. ↑ (⇑) Depicts an s-shell electron (hole) and a bold ⇑ marks p-shell hole.

To get the full picture of the specific few-particle complexes of charged exciton states (trion), we conducted micro-photoluminescence excitation spectroscopy (μPLE) measurements on the dominant X+ peak of both QD. The μPLE spectra of both QDs show a group of four resonances labeled as 1-4 in figure 2(a-b). Because the *hot* trion (X+*) consists of an e-h pair in the s-shell and a hole in the p-shell, there are $2^3$ possible spin configurations. According to Kramer's theorem all states have to be twofold degenerate without magnetic field, thus leaving only four degenerate doublets considering the h-h and e-h exchange interaction[1,2]. The four possible spin configurations of *hot* trion state are listed in the Fig. S2(c). The four resonances we found in the PLE experiments confirm the transitions path of 1-4 (indicated as purple arrow). Considering this and other evidences, we attribute the dominant μPL peaks to the radiative recombination from X+ state to the ground state.

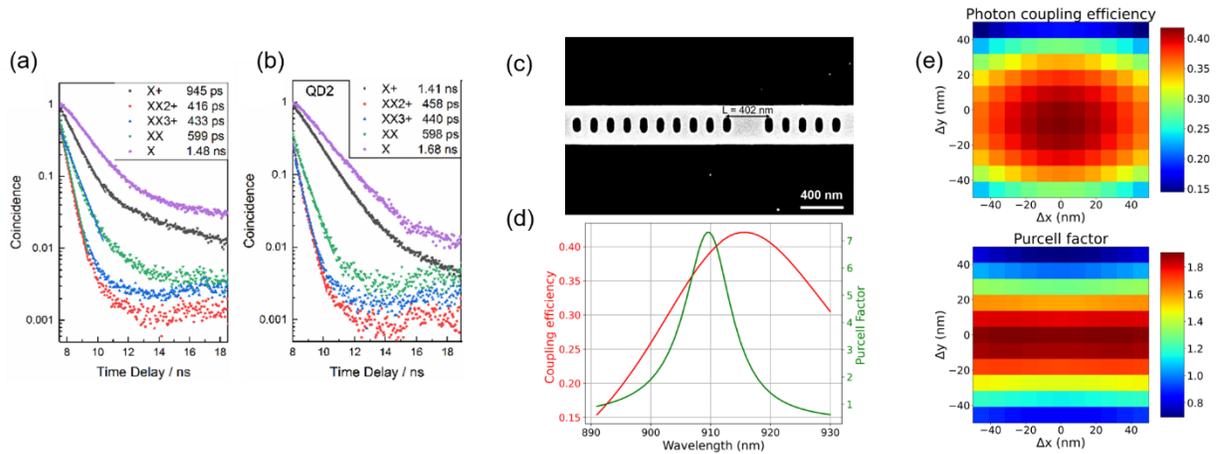

Fig. S3: Time resolved μPL measurements of integrated QDs. (a-b) Excitonic fluorescence decay traces of QD1 and QD2 at 4.3 K respectively. (c) Scanning electron microscope image of the fabricated nanobeam waveguide structure. (d) Simulated QD-WG coupling efficiency and Purcell factor versus wavelength for $L = 402$ nm of nanobeam cavity length. (e) 2D scan map of the calculated QD-waveguide coupling efficiency (upper) and Purcell factor (bottom) under lateral

displacement of the emitter with respect to the ideal position in the center of the fabricated nanobeam structure shown in (d) ($\lambda$ = 918 nm of dipole wavelength).

**(ii). Evaluation of Purcell enhancement for fabricated structures**

The Purcell enhancement of two integrated QDs was investigated by performing time-resolved µPL measurements for both QD-cavity structures. Fig. S3(a-b) illustrate respectively fluorescence decays traces of excitonic states for QD1 and QD2. At the temperature of 4.7 K, the fluorescence decays trace of $X^+$ state transition exhibits a lifetime of $\tau$ = (0.945±0.005) ns for QD1 and $\tau$ = (1.41±0.01) ns for QD2. No significant Purcell enhancement is observed compared with the lifetime of the same state transition of the planar QD ($\tau_{planar}$ = 1.57 ns, not shown here). We attribute the weak Purcell enhancement to the imperfect fabrication of the cavity. The cavity length of both nanobeam structures is 402 nm, which is 10 nm shorter than optimal length of 412 nm [See Fig. S3(c)]. To further confirm this assumption, the full 3D simulations were performed based on obtained device geometry. As illustrated in Fig. S3(d), The maximum Purcell factor is observed with a wavelength of around 910 nm, which is roughly a 10 nm blue shift. Moreover, the acquirable Purcell factor with 918 nm drops to 1.89. Furthermore, we combined the imperfect spatial displacement of the emitter in the cavity and this imperfectly-matched spectral alignment by implementing 2D scans of QD-WG coupling efficiency and Purcell factor as the function of displacements of QD ($\lambda$ = 918 nm, to fit emission of both QDs) inside the cavity [Fig. S3(e)]. The highest Purcell factor in obtained results is ~ 1.9. The non-negligible difference of 0.945 ns and 1.41 ns between the lifetimes of QD1 and QD2 on the same $X^+$ state, could be attributed to the QD1 being closer than the QD2 to the cavity mode.

**(iii). CW HOM measurements results.**

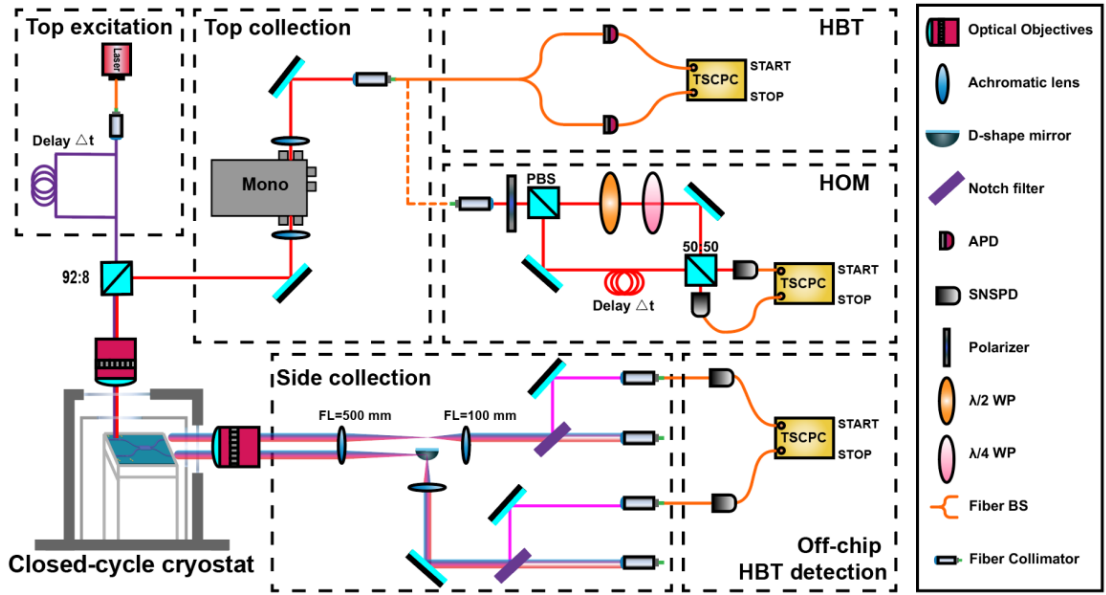

Figure S4. Schematic sketch of the used optical measurements setup including an off-chip HBT and HOM section, and an on-chip HBT section.

Optical measurements setup details are shown in Fig. S4. The setup includes a top excitation (and collection) path for initial characterizations and a side collection path for off-chip HBT detection experiments. An asymmetric Mach-Zehnder interferometer is inserted into the top excitation path to introduce a 4 ns time delay. It allows different choices on excitation laser type (ps pulsed laser or CW laser) since it is based on fiber. Conforming to the excitation path, a 4 ns time delay is also introduced in the detection path for HOM setup. For initial single-photon purity and lifetime evaluation, we used a monochromator in the top collection path to filter each excitonic state to perform time-correlation measurements. In the side collection path, two outputs emitting from cleaved waveguide ports facet were firstly collected by 50x optical objectives (NA = 0.42) and then separated by telescope system and D-shaped mirror. It also offers the option to collect the waveguide outputs in free-space and send them into a monochromator for splitting ratio and DOP measurements. Furthermore, for on-chip HBT experiments, we filtered the $X^+$ state by using a notch filter working in reflection mode.

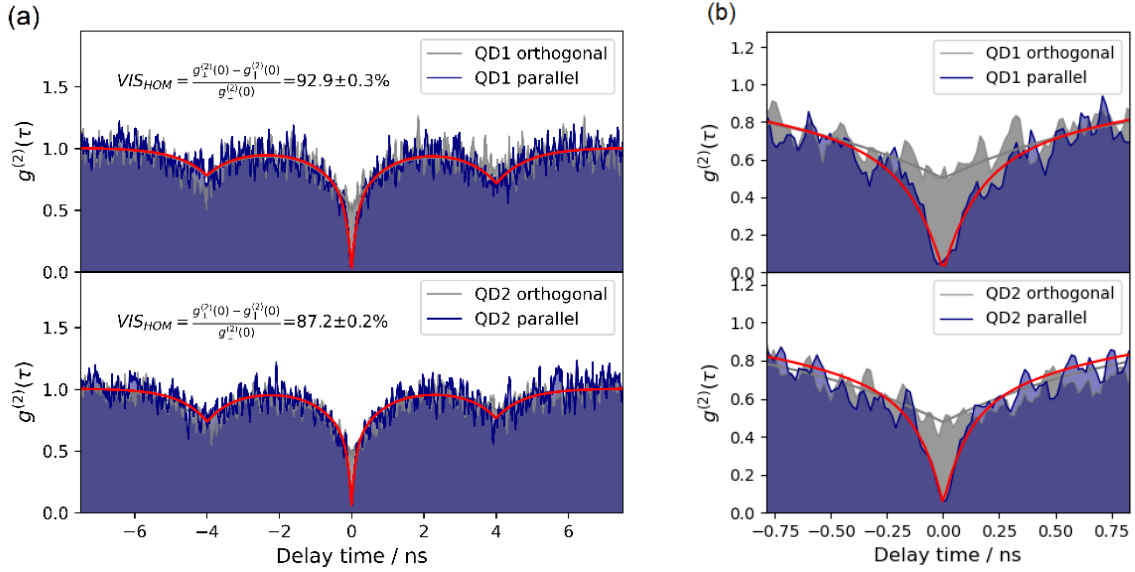

Fig. S5. Studying the indistinguishability of photons emitted by QD1 and QD2 deterministically integrated in the nanobeam waveguide structures. (a) Hong-Ou-Mandel interference measurements under continuous-wave excitation with 4 ns detection time-delay, revealing a post-selected TPI visibility of (92.9±0.3)% and (87.2±0.2)% for QD1 and QD2, respectively. (b) Zoom-in view of the correlation diagram with time window of 2 ns near zero-time delay.

To further verify the pulsed HOM results, we also conducted the HOM measurements upon aforementioned QDs under CW excitation condition. The results of two-photon interference experiments of both parallel and orthogonal polarization are shown in Fig. S5. Under orthogonal condition, $g^{(2)}_{HOM,\perp}(0) = 0.5$ is expected, and $g^{(2)}_{HOM,\parallel}(0)$ close to zero is expected under parallel polarizations whereas the photons are indistinguishable. Additionally, we observed two nearly equal dips ~0.75 near ±4 ns time delay, depicting near-to-ideal balanced beam splitter. Comparing the experimentally obtained $g^{(2)}_{\perp,\parallel}(0)$ using the function of $V_{HOM} = \frac{g^{(2)}_{HOM,\perp}(0) - g^{(2)}_{HOM,\parallel}(0)}{g^{(2)}_{HOM,\perp}(0)}$ yields a non-postselected HOM visibility of (92.9±0.3)% and (87.2±0.2)% for QD1 and QD2. To further compromise the multi-photon emission effect, the fitted value of HOM visibility is derived via the fitting function[3] as below:

$$g^{(2)}(t) = 1 - (1-A)\exp\left(-\left|\frac{t}{\tau_1}\right|\right)$$

$$g^{(2)}_{HOM,\perp}(t) = \overbrace{4(T_1^2 + R_1^2)T_2R_2 g^{(2)}(t)}^{G_1(t)} + \overbrace{4T_1R_1[T_2^2 g^{(2)}(t-\Delta t) + R_2^2 g^{(2)}(t+\Delta t)]}^{G_2(t)}$$

$$g^{(2)}_{HOM,\parallel}(t) = G_1(t) + G_2(t)\left[1 - V\exp\left(-\left|\frac{2t}{\tau_c}\right|\right)\right]$$

Here, $g^{(2)}(t)$ describes the anti-bunching behavior in the autocorrelation curve. $R_1 = 0.50$ and $T_1 = 0.50$ as well as $R_2 = 0.53$ and $T_2 = 0.47$ are the reflection and transmission indexes of the first and second beam splitter in the Mach−Zehnder interferometer in HOM setup. Fitting the orthogonal configuration of $g^{(2)}_{HOM,\perp}(t)$ in Fig. S5(a), we can extract $g^{(2)}(t)$. Then we use $g^{(2)}(t)$ to fit $g^{(2)}_{HOM,\parallel}(t)$ to the parallel polarization results with only $\tau_c$ and $V$ as free parameters. Thus, it leads to the fitted HOM visibility of (99.0 ± 0.8)% and (90.9 ± 0.7)% for QD1 and QD2. This large visibility value of more than 90% demonstrates a high value of post-selective indistinguishability of photons and a near optimal overlap of wave functions within their coherence time.

## (iv). Evaluation of the overall alignment precision of deterministic fabrication process.

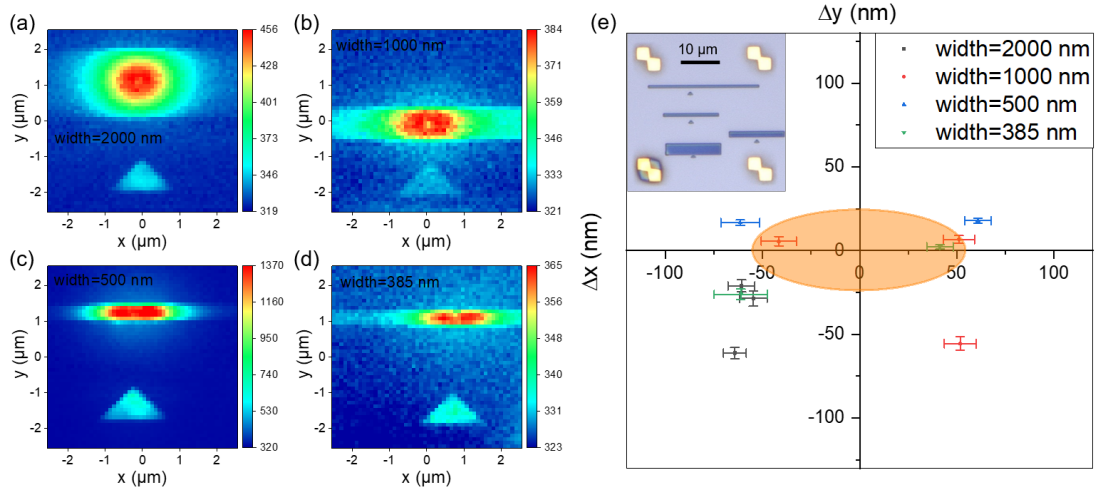

Fig. S6. Evaluation of the overall alignment accuracy between the centers of QDs and test waveguide. (a-d) Post CL maps of QD-waveguides test structures (with four different widths of 385 nm, 500 nm, 1000 nm, 2000 nm) to assess the overall accuracy of deterministic QD integration. (e) Integration errors of 10 test structures yielding an overall alignment accuracy of $\overline{|\Delta x|} = (54.8 \pm 7.9)$ nm, $\overline{|\Delta y|} = (24.2 \pm 18.9)$ nm, inset: optical microscope image of the test structures.

The overall fabrication precision includes four parts of alignment errors: stage drift (~5 nm/min) during the CL mapping, QD position extraction accuracy (2D gaussian fitting method gives ~5-7 nm precision), marker recognition accuracy (~30 nm) and EBL operation precision (~30 nm). To compensate for the stage drift error, we developed a homemade algorithm script to calibrate the CL map. In this script, we extract the center of the four markers in the corners of the writing field to obtain a transform matrix between the tilted field and the actual field. We then use the transform matrix to resize the four markers' coordinates and recalculate the coordinates of the QDs accordingly concerning the tilt angle and zoom factor. In this way, the stage drift error is almost eliminated since it only occurs in low-temperature CL maps capture and will not affect the room-temperature EBL fabrication process. In the EBL operation process, we from the alignment capabilities of our EBL system equipped with a high precision laser interferometer stage and an automized marker recognition program.

To evaluate the overall alignment accuracy between centers of QDs and the center of the nanostructure during the fabrication process, we performed test round of integrating QDs into ridge waveguides with four different widths (385 nm, 500 nm, 1000 nm, 2000 nm) under the same processing parameters and operation workflow as uses for the target structures of this work. The overall alignment error is defined by the difference between the target position and QD position. For the identification of the target position, the horizontal target is depicted by the triangle marker and the vertical target is given by the center of ridge waveguide. The precise QD position is extracted by a 2D Gaussian fitting method. We fabricated 10 test structures with pre-selection and post CL check, the integration successful rate is 10/10, and we obtained the standard deviation of the overall integration precision: $\overline{|\Delta x|} = (54.8 \pm 7.9)$ nm, $\overline{|\Delta y|} = (24.2 \pm 18.9)$ nm, which are depicted by an orange ellipse in Fig. S6 (e) and leading to the overall alignment accuracy of $\frac{\sqrt{|\Delta x|^2 + |\Delta y|^2}}{2} = (42 \pm 14)$ nm.